\documentclass[conference]{IEEEtran}
\IEEEoverridecommandlockouts
\usepackage{cite}
\usepackage{amsmath,amssymb,amsfonts}
\usepackage{algorithmic}
\usepackage{graphicx}
\usepackage{textcomp}
\usepackage{xcolor}
\usepackage{hyperref}
\usepackage{todonotes}
\usepackage{pdfpages}
\usepackage{fancyhdr}
\newcommand{\revise}[1]{\textcolor{black}{#1}}

\def\BibTeX{{\rm B\kern-.05em{\sc i\kern-.025em b}\kern-.08em
    T\kern-.1667em\lower.7ex\hbox{E}\kern-.125emX}}

\begin{document}

\title{Domain-Driven Design in Practice: A Mining Study of Maintenance and Evolution in Open-Source Repositories\\
\thanks{This work is funded by the German Federal Ministry for Research, Technology and Space (BMFTR) under grant number 01IS25017B (JDomInO project).}
}

\author{\IEEEauthorblockN{1\textsuperscript{st} Weixing Zhang}
\IEEEauthorblockA{
\textit{Karlsruhe Institute of Technology}\\
Karlsruhe, Germany \\
0000-0003-2890-6034}
\and
\IEEEauthorblockN{2\textsuperscript{nd} Bowen Jiang}
\IEEEauthorblockA{
\textit{Karlsruhe Institute of Technology}\\
Karlsruhe, Germany \\
0009-0001-8168-0866}
\and
\IEEEauthorblockN{3\textsuperscript{rd} Yuhong Fu}
\IEEEauthorblockA{
\textit{Adelaide University}\\
Adelaide, Australia \\
0000-0003-2093-2326}
\and
\IEEEauthorblockN{4\textsuperscript{th} Haowei Cheng}
\IEEEauthorblockA{
\textit{Faculty of Science and Engineering}\\
\textit{Waseda University}\\
Tokyo, Japan \\
0009-0008-2265-6437}
\and
\IEEEauthorblockN{5\textsuperscript{th} Mario Herb}
\IEEEauthorblockA{
\textit{esentri AG}\\
Karlsruhe, Germany \\
0009-0008-9423-4786}
\and
\IEEEauthorblockN{6\textsuperscript{th} Anne Koziolek}
\IEEEauthorblockA{
\textit{Karlsruhe Institute of Technology}\\
Karlsruhe, Germany \\
0000-0002-1593-3394}
}

\maketitle
\thispagestyle{fancy}

\begin{abstract}
Domain-Driven Design (DDD) is an influential software development methodology that structures software around business domain complexity through tactical building blocks such as Entities, Value Objects, Aggregates, and Repositories. Despite its prominence in software engineering, large-scale empirical evidence on how DDD is practiced, how it evolves, and how it relates to software maintenance quality in open-source projects remains scarce. 
This study presents a pre-registered empirical investigation of the distribution, evolution, and maintenance implications of DDD tactical building blocks in open-source GitHub repositories, addressing the call for large-scale empirical evaluations identified in a recent systematic literature review. We will collect DDD-related repositories from GitHub using the Search API, apply automated keyword filtering and manual relevance assessment, and analyze the resulting dataset through four research questions covering: (RQ1) the static distribution and co-usage of DDD building blocks across repository types, (RQ2) their longitudinal evolution over commit history, (RQ3) the extent and maintenance implications of Bounded Context boundary violations (the primary technical challenge in DDD adoption, unquantified at scale), and (RQ4) the temporal association between maintenance activities and building block churn or Bounded Context violations in DDD repositories.
Together, these RQs trace the full maintenance and evolution lifecycle of open-source DDD projects and establish an empirical foundation for future DDD tool support and methodology refinement.
\end{abstract}


\section{Introduction}
Domain-Driven Design (DDD), introduced by Evans~\cite{evans2004domain}, is a software development methodology that aligns software structure with the complexity of the business domain. By organizing software around a rich domain model expressed through tactical building blocks, e.g., Entities, Value Objects, Aggregates, Repositories, Domain Services, and Domain Events, DDD has become a prominent approach to managing complexity in large-scale software systems. Its core concepts, including the notion of Bounded Contexts and Ubiquitous Language, have gained traction in both industry and open-source development over the past two decades~\cite{evans2004domain}\cite{vernon2013implementing}.

Despite its widespread adoption, the empirical foundation of DDD remains thin. Existing studies rely predominantly on case studies, expert opinions, or theoretical arguments, with very few large-scale empirical validations (Özkan et al.~\cite{ozkan2025ddd}). At the practitioner level, an evidence-based investigation combining a systematic literature review and an industrial survey of 63 practitioners confirms that the gap between domain model design and its code implementation is a recurring challenge in real-world projects, affecting teams regardless of their experience level (Zhong et al.~\cite{zhong2024domain}). Both studies converge on the same root cause: the divergence between the intended domain model and its actual implementation, manifested in practice as Bounded Context boundary violations, remains the primary technical challenge in DDD adoption, yet its prevalence and consequences in open-source projects have not been systematically quantified at scale. 
This gap between DDD's recognized significance in software engineering discourse and the scarcity of large-scale empirical evidence motivates the present study~\cite{zhang2023manual}, \cite{zhang2025empirical}.

The maintenance and evolution of DDD repositories are particularly underexplored. While DDD is explicitly motivated by the need to manage long-term software complexity, little is known about how DDD tactical building blocks evolve as projects mature, whether Bounded Context boundaries are preserved over time, or how structural degradation relates to observable maintenance activity. Addressing these questions requires evidence at a scale that case studies and practitioner surveys cannot provide. Open-source repositories hosted on GitHub provide a valuable opportunity for this kind of large-scale observation: a preliminary search querying repositories tagged with ``domain-driven-design'' and ``ddd'' with a minimum threshold of ten stars identified 1,260 repositories; after automated filtering, 865 remained, confirming that a sufficient corpus exists to support large-scale analysis. Mining this corpus enables observations of DDD as it is actually practiced across diverse projects, languages, team sizes, and repository purposes, rather than as it is prescribed in textbooks or reported in selected case studies. These questions are central to the scope of the International Conference on Software Maintenance and Evolution (ICSME), and the present study is directly positioned to contribute empirical evidence to this theme.



This study is conducted in the context of the JDomInO project, a funded academic-industry collaboration between Karlsruhe Institute of Technology (KIT) and esentri AG that investigates the integration of Domain-Driven Design with Model-Driven Engineering (MDE) to support the development of domain-specific toolchains~\cite{zhang2026round}.
The empirical findings of the present study are intended to inform the theoretical and tooling work of JDomInO by providing large-scale evidence on how DDD is practiced in the wild, where the model-code gap manifests, and what maintenance patterns are associated with DDD adoption. In particular, the prevalence and consequences of BC boundary violations identified in this study will directly inform the constraint rule design in JDomInO's modeling toolchain.

\revise{BC boundary violation detection is conceptually related to several established research traditions. Architecture erosion research has characterized the progressive gap between intended and implemented architectures across general software systems~\cite{li2022understanding}, and architecture conformance checking approaches have operationalized this gap through static dependency analysis~\cite{maffort2016mining}. Architectural smell detection, including microservice-specific smells such as cyclic dependencies and shared persistence violations~\cite{mumtaz2021systematic}, addresses structurally similar phenomena at a finer granularity. On the tooling side, Context Mapper~\cite{kapferer2020domain} and ArchUnit~\cite{archunit} provide practitioners with rule-based BC boundary enforcement, but without large-scale empirical evidence on how such boundaries evolve or erode in the wild.}

\revise{By operationalizing BC boundary violations in the DDD-specific context and measuring their prevalence and maintenance implications at scale, a scope not addressed by prior architecture conformance or smell detection work, the present study makes the following contributions.}
First, it provides the first large-scale empirical baseline of DDD tactical building block usage across open-source repositories spanning multiple programming languages. Second, it characterizes the longitudinal evolution of DDD building block usage over commit history. Third, it operationalizes and measures Bounded Context boundary violations at scale and examines their association with maintenance quality. Fourth, it analyzes maintenance activity patterns in DDD repositories and their temporal association with structural change. Together, these contributions address the call in Özkan et al.~\cite{ozkan2025ddd} for systematic large-scale empirical evaluations of DDD in practice, and establish an empirical foundation for future work on DDD tool support and methodology refinement.

\section{Research Questions}
This is an Exploratory study; we therefore formulate research questions rather than fixed hypotheses. The four RQs together trace the maintenance and evolution lifecycle of open-source DDD projects, from static usage through evolutionary dynamics to maintenance challenges and developer responses.

\textbf{RQ1:} How are DDD tactical building blocks distributed and co-used in open-source repositories?

This RQ establishes the empirical baseline of the study. Beyond simple frequency counts, we investigate the structural co-occurrence relationships between building block types and whether usage completeness correlates with project-level characteristics such as maturity, size, and community engagement.
Building block usage is further analyzed stratified by repository type, including \textit{Production Application}, \textit{Framework/Library}, and \textit{DDD Tooling/Infrastructure} (see~\ref{sec:confounding}).
Cross-language comparisons reveal whether DDD practice is shaped by language ecosystem affordances. This RQ directly addresses the call in Ozkan et al.~\cite{ozkan2025ddd} for large-scale empirical evaluations of DDD in practice, as their review found that the majority of existing studies rely on case studies or theoretical arguments without systematic large-scale validation.

\textbf{RQ2:} How do DDD tactical building block usage evolve over the commit history of open-source projects?

This RQ shifts from a static snapshot to a longitudinal perspective, examining how building block usage changes as projects mature. We track the temporal introduction and removal of building blocks, and whether their proportional composition shifts over time. 
Large-scale restructuring commits are identified and treated separately to isolate incremental drift from deliberate architectural reorganization. 

\textbf{RQ3:} To what extent are Bounded Context boundary violations observable in open-source DDD repositories, and how do they associate with maintenance quality?

The model-code gap, operationalized here as the erosion of Bounded Context boundaries through cross-context dependencies, is identified as the primary technical challenge in DDD adoption~\cite{ozkan2025ddd}. This RQ is scoped to this single, tractable manifestation of the model-code gap rather than attempting a comprehensive gap measurement, in order to keep the analysis computationally feasible and the results interpretable. 
The rationale for this scoping decision, the operationalization of BC boundary inference and violation detection, and the maintenance quality proxies used in the association analysis are detailed in Sections~\ref{sec:variables} and~\ref{sec:execution_plan}, respectively.

\textbf{RQ4:} What maintenance activities are observable in DDD repositories, and do they temporally associate with building block evolution and Bounded Context violations?

This RQ examines the response side of the maintenance lifecycle. Commit messages are classified into maintenance activity types, and the central analysis is a temporal association analysis: whether DDD-specific restructuring commits cluster following periods of elevated BC violations from RQ3 or rapid building-block churn from RQ2. 
The classification protocol and sliding-window-based temporal analysis method are detailed in Section~\ref{sec:execution_plan}.

\revise{If the study succeeds, these four RQs collectively establish the first large-scale empirical baseline of DDD practice in open-source repositories. If key detection steps fail to meet precision thresholds, the pre-defined degradation plans detailed in Section~\ref{sec:execution_plan} ensure that the scope of conclusions is narrowed accordingly, rather than invalidating the study as a whole.}

\section{Variables}
\label{sec:variables}

This study does not test pre-specified hypotheses. Variables are presented as constructs and their operationalizations, as our research questions concern descriptive distributions, evolutionary trajectories, and associations rather than causal relationships.

\subsection{Primary Constructs and Operationalizations}\label{AA}
\label{sec:primary_construct}
\begin{figure}[htbp]
    \centering
    \includegraphics[width=\columnwidth]{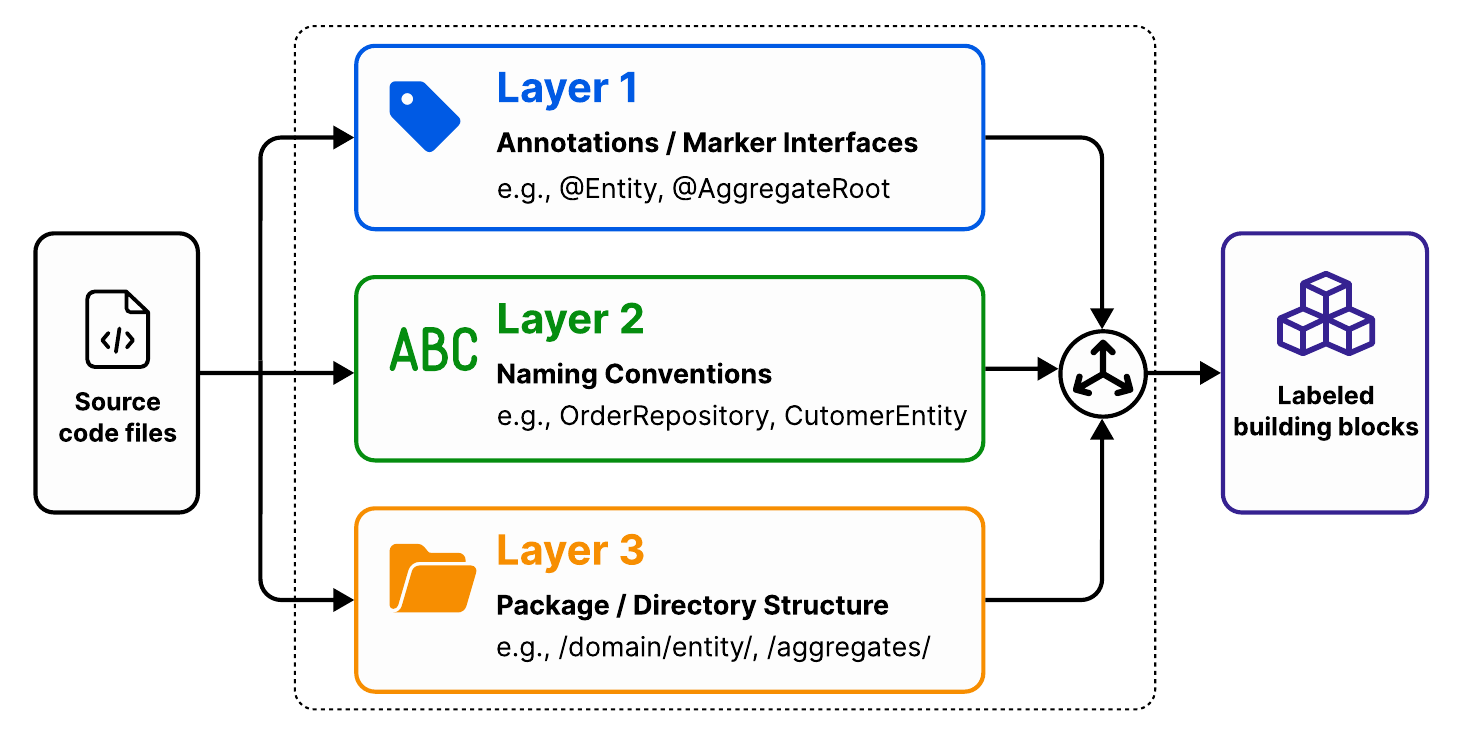}
    \caption{Three-layer identification pipeline for DDD tactical building blocks.}
    \label{fig:pipeline}
\end{figure}

\paragraph{DDD Tactical Building Block} A class identified as one of \texttt{Entity}, \texttt{Value Object}, \texttt{Aggregate Root}, \texttt{Repository}, \texttt{Domain Service}, \texttt{Domain Event}, \texttt{Application Service}, or \texttt{Factory}. Identification proceeds via a three-layer pipeline: (1) DDD-specific annotations or marker interfaces such as JMolecules~\cite{jmolecules_github} @Entity or domainlifecycles annotations; 
(2) 
naming conventions where the class name contains a building block type keyword, for example \texttt{OrderRepository} or \texttt{CustomerEntity}; however, such conventions are also prevalent in non-DDD contexts, most notably in projects using JPA or Spring Data~\cite{spring_repository_javadoc}. To quantify this risk, Phase 2 validation (Section~\ref{sec:phase_2}) will report precision separately for Layer 2-only detections; if precision falls below 0.75, such detections will be flagged as low-confidence and excluded from the main analysis.
\revise{The validation procedure and precision threshold for this inference method are detailed in Section~\ref{sec:phase_5}.}
(3) package or directory structure where the class resides in a package named after a building block type, such as /domain/entity/ or /aggregates/. 
Results from higher layers take precedence over lower layers in case of conflict. The applicability of Layer 1 is language-dependent; for languages lacking established DDD annotation ecosystems (e.g., TypeScript, PHP), identification relies primarily on Layers 2 and 3. The overall pipeline is illustrated in Figure~\ref{fig:pipeline}.
\revise{False positives are controlled at each layer; the corresponding validation thresholds and fallback procedures are detailed in Sections~\ref{sec:phase_2} and~\ref{sec:phase_5}.}

\paragraph{Building Block Usage Pattern} The multiset of building block types present in a repository at a given commit snapshot, characterized by their counts and proportional distribution.

\paragraph{Building Block Evolution} Change in building block usage across commits. Measured as: (a) absolute count delta per building block type between consecutive quarterly snapshots; (b) distributional shift measured by Jensen-Shannon divergence between consecutive quarterly snapshots.


\paragraph{Bounded Context Boundary} Inferred from top-level domain package and module structure and directory naming conventions. This operationalization assumes that BC boundaries are reflected in the package or module structure of the repository; projects where BC boundaries are expressed through inline comments, external documentation, or runtime configuration are outside the scope of this inference method and will be excluded from \textbf{RQ3} analysis if detected during manual validation. 
Monorepos and multi-module Maven/Gradle projects are handled at the module level where applicable; cases where reliable inference remains impossible are flagged and excluded.
Validated through two-author independent manual annotation on a stratified sample of 60 repositories, with Cohen's kappa $\geq$ 0.80 required and a minimum precision of 0.75 required before large-scale application.

\paragraph{BC Boundary Violation} An import or dependency from a class in one inferred BC to a class in a different inferred BC, detected via static import analysis. Violation rate is computed as the number of cross-BC dependencies divided by the total inter-class dependencies in the repository.

\paragraph{Maintenance Activity Type} Classified from commit messages into Corrective, Perfective, Adaptive, and DDD-Specific Restructuring as detailed in~\ref{sec:phase_6}, Phase 6.

\paragraph{Maintenance Quality Proxies} Bug-fixing commit frequency measured as corrective commits per month, commit activity recency measured as days since last commit, open issue count, and code churn rate measured as lines added plus deleted per commit, averaged over the past six months. \revise{These proxies are indirect and potentially noisy measures of maintenance quality rather than direct assessments of maintainability.}

\subsection{Confounding Factors}
\label{sec:confounding}
We identify the following confounding factors and describe how they will be considered in the analysis: 1) Project age and size: reported as covariates in all correlation analyses; stratified analysis performed where appropriate. 2) Primary programming language: results for RQ2 and RQ3 reported per primary language for languages with at least 100 repositories; no aggregate cross-language statistics presented for these RQs due to tool heterogeneity. \revise{The primary language of each repository will be determined using the GitHub API language field, which reports the most prevalent language by bytes of code. For repositories where this field is unavailable or inconclusive, primary language will be determined through manual inspection of the repository's source file composition.} 3) Repository type: GitHub repositories serve highly heterogeneous purposes, and conflating structurally distinct repository types can introduce systematic bias into the analysis (Kalliamvakou et al.~\cite{kalliamvakou2014promises}). Following the precedent of technology-specific repository studies ~\cite{zhang2024tales}\cite{zhang2026development}, we adopt a DDD-tailored classification rather than Kalliamvakou et al.'s general-purpose taxonomy. We therefore distinguish three types: \textit{Production Application} (repositories implementing domain logic for a real business domain), \textit{Framework/Library} (repositories providing reusable DDD-related abstractions for consumption by other projects), and \textit{DDD Tooling/Infrastructure} (repositories providing tooling support for DDD practice itself (e.g., code generators, annotation processors, or scaffolding tools) without encoding business domain logic). The third category represents a population largely absent from general GitHub studies, motivated by the expectation that a share of DDD-tagged repositories may serve infrastructural rather than domain-modeling purposes. Classification is performed manually during the relevance filtering stage using two-author independent coding; inter-rater reliability is validated via Cohen's kappa, targeting $\kappa$ $\geq$ 0.80. Repository type is used as a grouping variable in RQ1 and RQ3. 4) Team size: approximated by the number of distinct committers; reported as a covariate.

\section{Participants / Subjects / Datasets}

\subsection{Repository Collection}
\label{sec:repo_collection}
We will collect DDD-related repositories from GitHub using the GitHub Search API. Queries will target repositories tagged with the topics domain-driven-design and ddd, supplemented by keyword searches in repository names and descriptions. No minimum star threshold and no activity threshold will be imposed at collection time, in order to achieve exhaustive coverage and avoid excluding historically valuable but currently inactive repositories. After merging and deduplicating results, an automated relevance pre-filter will be applied: repositories whose description and README contain none of the following keywords: domain-driven design, domain driven design, ddd, bounded context, aggregate root, ubiquitous language, will be excluded. Remaining repositories will proceed to the two-stage manual relevance filtering described below.

\subsection{Repository Relevance Filtering}
The automated filter applies two checks. 
First, relevance to DDD is assessed by verifying that the repository's topic tags, description, or README contains at least one of the following terms: domain-driven design, domain driven design, ddd, bounded context, aggregate root, ubiquitous language. These terms were selected as the highest-level signals of DDD as a core methodology: the first two are the methodology name itself, and the remaining four represent the most distinctive strategic and tactical concepts of DDD. More fine-grained DDD terms such as Domain Event or Value Object were deliberately excluded, as they may appear in non-DDD contexts and would reduce filtering precision. 
Second, the repository must contain at least one non-empty source code file, verified via the GitHub API file tree. 
Third, the repository is not a fork of another repository, verified via the GitHub API fork field. However, since repositories that were originally forked but have since diverged from their upstream source may represent genuine independent projects, we apply an additional criterion: 
forked repositories will be re-included if they satisfy both of the following conditions: (a) the fork was created more than 24 months prior to data collection, and (b) the number of commits unique to the fork exceeds the number of commits inherited from the upstream repository. 

To estimate the precision of the automated filter, a stratified random sample of the surviving repositories is drawn, and both authors independently assess the same sampled repositories, without knowledge of each other's decisions. Each repository is evaluated to determine whether it genuinely practices DDD as a design methodology. The following exclusion criteria serve as the ground truth definition for this assessment: a sampled repository is counted as a false positive of the automated filter if it meets any of the criteria below. Precision is then computed as the proportion of sampled repositories that meet none of the exclusion criteria.
Cohen's kappa will be computed to validate inter-rater reliability, targeting kappa $\geq$ 0.80. Disagreements are resolved through structured discussion; a third author acts as a tiebreaker if needed. The estimated precision will be reported as an indicator of dataset quality. 
If precision falls below 0.70, we will analyze the false positive cases to identify systematic patterns, refine the automated filter accordingly, and reapply it to the full collection before proceeding.

Exclusion criteria:
\begin{itemize}
    \item The repository only mentions DDD or uses it as a dependency, without genuinely applying it as a design methodology.
    \item The repository's DDD content is limited to example code accompanying a book or tutorial, with no original domain logic.
\end{itemize}

It should be noted that the relevance filtering criteria above cannot distinguish repositories that apply DDD tactical building blocks from those that adopt DDD terminology without implementing the underlying building block structure. The findings of this study therefore characterize DDD tactical building block usage as observed in self-identified DDD repositories, not as a validation of correct DDD practice.
\revise{In other words, this study measures code-level DDD signals rather than true DDD design quality, and claims in the Stage 2 paper will be scoped accordingly. It should be noted that our fork exclusion criterion relies on GitHub API forge fork metadata, and therefore does not account for repositories forked exogenously to the platform via git clone or similar workflows~\cite{pietri2020forking}; such Type 2 forks, if present in the dataset, represent a potential source of duplication that we acknowledge as a limitation.}

\subsection{Repository Maintenance Status}
For the purpose of this study, we operationalize repository maintenance status into three categories. Active means at least one commit in the past 12 months. Maintained (low activity) means no commits in the past 12 months but at least one commit in the past 36 months. Inactive means no commit in the past 36 months. These categories will be used as a grouping variable in RQ1 and RQ3, and reported as a descriptive characteristic of the dataset.

\subsection{Language Scope for Deep Analyses}
The target languages for deep analyses (RQ2 and RQ3) are Java, C\#, and TypeScript. These three languages are selected on the basis of the preliminary corpus: among the 865 repositories identified in the feasibility search, Java, C\#, and TypeScript collectively account for the majority of repositories, and each is supported by an AST-level parsing ecosystem. Languages represented by fewer than 100 repositories in the final dataset will be excluded from RQ2 and RQ3 deep analyses and included only in the language-agnostic RQ1 and RQ4 analyses. The exact language distribution of the final dataset will be reported in the Stage 2 paper. RQ2 requires source file parsing to support the building block identification pipeline (Phases 1–2 of the pipeline described in Section~\ref{sec:primary_construct}); for naming convention and package structure analysis, text-based processing suffices, while annotation detection may require lightweight parsing depending on the target language. RQ3 requires AST-level static analysis to accurately resolve cross-BC import dependencies, for which language-native parsers are used (JavaParser~\cite{javaparser_github} for Java, Roslyn~\cite{roslyn_github} for C\#, TypeScript Compiler API~\cite{typescript_compiler_api} for TypeScript).

\subsection{Dataset Feasibility and Expected Scale}
\label{sec:data_scale}
To assess the feasibility of the study, we conducted a preliminary search using the GitHub Search API, querying repositories tagged with the topics ``domain-driven-design'' and ``ddd'', with a minimum threshold of 10 stars, yielding 1,260 repositories after merging and deduplication; after automated filtering, 865 remained.
The present study removes the star threshold, and the initial collection is therefore expected to be larger. These results confirm that a sufficient number of DDD repositories exists on GitHub to support the planned large-scale analyses.

\section{Execution Plan}
\label{sec:execution_plan}
\begin{figure*}[t]
    \centering
    \includegraphics[width=\linewidth]{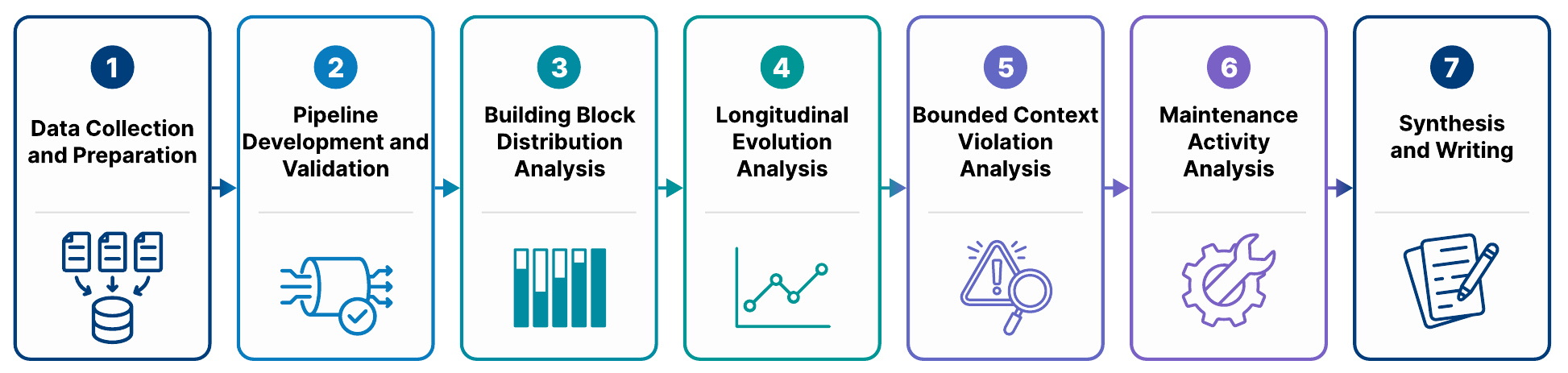}
    \caption{Seven-phase execution plan of the study.}
    \label{fig:execution_plan}
\end{figure*}

This section describes the seven-phase execution plan of the study, as illustrated in Figure~\ref{fig:execution_plan}.

\subsection{Phase 1: Data Collection and Preparation}
In Phase 1, we perform the following tasks: 1) Execute GitHub Search API queries for exhaustive repository collection; deduplicate and merge results. 2) Apply automated relevance filter (i.e., automated keyword matching). 3) Conduct two-author independent manual relevance assessment; compute Cohen's kappa per batch of 50 repositories; resolve disagreements via structured discussion with third-author tiebreaker if needed. 4) Classify each repository into one of three types, i.e., \textit{Production Application}, \textit{Framework/Library}, or \textit{DDD Tooling/Infrastructure}, as defined in Section III.B, using two-author independent coding; record maintenance status per the operationalization in Section~\ref{sec:confounding}. 5) Clone all final repositories locally; extract full commit histories using PyDriller~\cite{spadini2018pydriller}.

\subsection{Phase 2: Building Block Identification Pipeline Development and Validation}
\label{sec:phase_2}
In Phase 2, we perform the following tasks: 1) Implement Layer 1 (annotation and marker-interface detection) for each target language. 2) Implement Layer 2 (naming-convention regex matching) per language. 3) Implement Layer 3 (package and directory structure analysis) per language. 4) Apply pipeline to the latest commit snapshot of all final repositories. 5) Validate pipeline on a stratified random sample of 50 repositories via two-author manual inspection; compute and report precision and recall. A minimum precision of 0.75 is required before proceeding to large-scale analysis.

\subsection{Phase 3: RQ1 - Building Block Distribution Analysis}
In Phase 3, we perform the following tasks: 1) Compute frequency and co-occurrence statistics for all building block types across the full dataset. 2) Perform cross-language comparisons using Kruskal-Wallis tests~\cite{kruskal1952use} with Bonferroni-corrected~\cite{bland1995multiple} Dunn post-hoc pairwise tests~\cite{dunn1964multiple}. 3) Correlate building block usage completeness with project characteristics (age, size, stars, forks, maintenance status) using Spearman's rho~\cite{spearman1961proof}. 4) Analyze building block usage stratified by repository type.

\subsection{Phase 4: RQ2 - Longitudinal Evolution Analysis}
In Phase 4, we perform the following tasks: 1) Apply building block identification pipeline to commit snapshots at quarterly intervals for all target-language repositories. 2) Filter out commits affecting only non-source files (documentation, configuration, tests) using file-extension rules. 3) Large restructuring commits are identified and flagged to prevent conflating deliberate architectural reorganization with incremental drift. The threshold for classifying a commit as a large restructuring commit will be determined empirically: upon completion of data collection, we will compute the distribution of files changed per commit across all repositories and adopt the 95th percentile as the threshold. Sensitivity analyses at the 90th and 99th percentiles will be conducted to verify that the main findings are not sensitive to the threshold choice. Flagged commits will be excluded from the main trend analysis, and their frequency and size distribution will be reported separately. The files-changed distribution and resulting threshold will be reported in the Stage 2 paper and replication package. 4) Compute absolute count changes per building block type between consecutive quarterly snapshots. 5) Compute distributional shift using Jensen-Shannon divergence~\cite{lin2002divergence} between consecutive quarterly snapshots (see Section~\ref{sec:primary_construct}). As JSD thresholds are inherently domain-dependent, JSD values will be interpreted descriptively rather than against a fixed cutoff: repositories will be ranked by their JSD scores, and the distribution of JSD values across the dataset will be reported, with the 75th and 90th percentiles used as reference points to identify repositories exhibiting compositional drift. 6) Analyze temporal patterns: building block introduction sequences, stabilization points, and long-term compositional drift.

\subsection{Phase 5: RQ3 - Bounded Context Violation Analysis}
\label{sec:phase_5}
In Phase 5, we perform the following tasks: 1) Infer BC boundaries from package and module structure for each target-language repository. 2) Conduct two-author independent manual validation of inferred boundaries on a stratified sample of 60 repositories (20 per language); compute Cohen's kappa and precision; proceed only if precision $\geq$ 0.75. Disagreements resolved via structured discussion using a pre-defined decision protocol. If the precision threshold of 0.75 is not met for a given language, the RQ3 results for that language will be excluded from cross-language comparisons and reported separately as a limitation. If no language reaches the threshold, this null result will be reported as an empirical finding and RQ3 reframed accordingly. 3) Detect cross-BC import dependencies per the operationalization in Section~\ref{sec:primary_construct}. Compute BC violation rate per repository. 4) Report results per language without cross-language aggregation, due to heterogeneity in tool precision across languages. 5) Correlate BC violation rates with maintenance quality proxies using Spearman's rho, reported per language.

\subsection{Phase 6: RQ4 - Maintenance Activity Analysis}
\label{sec:phase_6}
In Phase 6, we perform the following tasks: 1) Finalize commit message classification taxonomy with keyword lists and decision rules, organized into four categories: Corrective (bug fixes), Perfective (refactoring and code quality), Adaptive (dependency updates and API changes)~\cite{lientz1980software}, and DDD-Specific Restructuring (commits explicitly referencing DDD concepts such as aggregate, bounded context, entity, or value object). The prevalence of DDD-specific terminology in commit messages is an empirical question that will be answered during execution. If fewer than 5\% of commits in the dataset are classified as DDD-Specific Restructuring, this category will be reported descriptively only and excluded from the temporal association analysis. 2) Apply keyword-based classifier to all commit messages across the full dataset. 3) Apply LLM-assisted classification to ambiguous cases, defined as commits that match keywords from more than one category or match no keywords at all. LLM confidence scores are used to triage the results: low-confidence classifications are subject to mandatory two-author manual review, while high-confidence classifications are subject to random spot-checking at a 10\% sampling rate. The LLM prompt, and triage decision rules are documented in the replication package. \revise{Together, these two tiers ensure that ambiguous cases are handled systematically rather than left to ad hoc judgment. Confidence thresholds will be determined in advance through a pilot experiment before large-scale application, and will be fully documented in the replication package.} 4) Validate the classification pipeline on a stratified random sample of 200 commits (50 per category), ensuring sufficient representation of each maintenance category (±14\% margin at 95\% CI, acceptable for validation purposes). Two-author independent coding is applied to all 200 commits, targeting Cohen's kappa $\geq$ 0.80. 5) Perform temporal association analysis using a 30-day sliding window, following prior MSR work that adopted this window size as capturing a meaningful unit of development activity while remaining smaller than typical release intervals~\cite{jiang2013mining}, to measure the lag between periods of elevated structural change (from RQ2 and RQ3) and subsequent maintenance activity spikes. Sensitivity analyses using 14-day and 60-day windows will be conducted to verify that findings are not sensitive to the window size choice. Spearman's rho is computed on the resulting time-series data to quantify the temporal association.

\subsection{Phase 7: Synthesis and Writing}
In Phase 7, we perform the following tasks: 1) Synthesize findings across all four RQs into a coherent narrative addressing the maintenance and evolution lifecycle of DDD projects; 2) Prepare and publish an open replication package (dataset, scripts, taxonomies, annotation guidelines, LLM prompts); 3) Write, internally review, and revise the Stage 2 manuscript for submission to EMSE.





\clearpage

\bibliographystyle{IEEEtran}
\bibliography{icsme26}

@article{ozkan2025ddd,
  author       = {Ozan {\"{O}}zkan and
                  {\"{O}}nder Babur and
                  Mark van den Brand},
  title        = {Domain-Driven Design in software development: {A} systematic literature
                  review on implementation, challenges, and effectiveness},
  journal      = {J. Syst. Softw.},
  volume       = {230},
  pages        = {112537},
  year         = {2025},
  url          = {https://doi.org/10.1016/j.jss.2025.112537},
  doi          = {10.1016/J.JSS.2025.112537},
  bibsource    = {dblp computer science bibliography, https://dblp.org}
}

@inproceedings{kalliamvakou2014promises,
  author       = {Eirini Kalliamvakou and
                  Georgios Gousios and
                  Kelly Blincoe and
                  Leif Singer and
                  Daniel M. Germ{\'{a}}n and
                  Daniela E. Damian},
  editor       = {Premkumar T. Devanbu and
                  Sung Kim and
                  Martin Pinzger},
  title        = {The promises and perils of mining GitHub},
  booktitle    = {11th Working Conference on Mining Software Repositories, {MSR} 2014,
                  Proceedings, May 31 - June 1, 2014, Hyderabad, India},
  pages        = {92--101},
  publisher    = {{ACM}},
  year         = {2014},
  url          = {https://doi.org/10.1145/2597073.2597074},
  doi          = {10.1145/2597073.2597074},
  bibsource    = {dblp computer science bibliography, https://dblp.org}
}

@inproceedings{spadini2018pydriller,
  author       = {Davide Spadini and
                  Maur{\'{\i}}cio Finavaro Aniche and
                  Alberto Bacchelli},
  editor       = {Gary T. Leavens and
                  Alessandro Garcia and
                  Corina S. Pasareanu},
  title        = {PyDriller: Python framework for mining software repositories},
  booktitle    = {Proceedings of the 2018 {ACM} Joint Meeting on European Software Engineering
                  Conference and Symposium on the Foundations of Software Engineering,
                  {ESEC/SIGSOFT} {FSE} 2018, Lake Buena Vista, FL, USA, November 04-09,
                  2018},
  pages        = {908--911},
  publisher    = {{ACM}},
  year         = {2018},
  url          = {https://doi.org/10.1145/3236024.3264598},
  doi          = {10.1145/3236024.3264598},
  bibsource    = {dblp computer science bibliography, https://dblp.org}
}

@article{kruskal1952use,
  title={Use of ranks in one-criterion variance analysis},
  author={Kruskal, William H and Wallis, W Allen},
  journal={Journal of the American statistical Association},
  volume={47},
  number={260},
  pages={583--621},
  year={1952},
  publisher={Taylor \& Francis}
}

@article{dunn1964multiple,
  title={Multiple comparisons using rank sums},
  author={Dunn, Olive Jean},
  journal={Technometrics},
  volume={6},
  number={3},
  pages={241--252},
  year={1964},
  publisher={Taylor \& Francis}
}

@article{bland1995multiple,
  title={Multiple significance tests: the Bonferroni method},
  author={Bland, J Martin and Altman, Douglas G},
  journal={Bmj},
  volume={310},
  number={6973},
  pages={170},
  year={1995},
  publisher={British Medical Journal Publishing Group}
}

@article{spearman1961proof,
  title={The proof and measurement of association between two things.},
  author={Spearman, Charles},
  year={1961},
  publisher={Appleton-Century-Crofts}
}

@article{lin2002divergence,
  title={Divergence measures based on the Shannon entropy},
  author={Lin, Jianhua},
  journal={IEEE Transactions on Information theory},
  volume={37},
  number={1},
  pages={145--151},
  year={2002},
  publisher={IEEE}
}

@book{lientz1980software,
  title={Software maintenance management},
  author={Lientz, Bennett P and Swanson, E Burton},
  year={1980},
  publisher={Addison-Wesley Longman Publishing Co., Inc.}
}

@article{jiang2013mining,
  title={Mining the temporal evolution of the android bug reporting community via sliding windows},
  author={Jiang, Feng and Wang, Jiemin and Hindle, Abram and Nascimento, Mario A},
  journal={arXiv preprint arXiv:1310.7469},
  year={2013}
}

@misc{spring_repository_javadoc,
  author       = {{VMware, Inc.}},
  title        = {{@Repository} Annotation --- {Spring Framework} 7.0.x {JavaDoc API}},
  year         = {2025},
  url          = {https://docs.spring.io/spring-framework/docs/current/javadoc-api/org/springframework/stereotype/Repository.html},
  note         = {Accessed: 2025-05-02}
}

@misc{jmolecules_github,
  author    = {{xMolecules Contributors}},
  title     = {{jMolecules}: Libraries to Help Developers Express Architectural Abstractions in {Java} Code},
  year      = {2021},
  url       = {https://github.com/xmolecules/jmolecules},
  note      = {Accessed: 2025-05-02}
}

@misc{javaparser_github,
  author    = {{JavaParser Contributors}},
  title     = {{JavaParser}: {Java} 1--21 Parser and Abstract Syntax Tree for {Java}},
  year      = {2013},
  url       = {https://github.com/javaparser/javaparser},
  note      = {Accessed: 2025-05-02}
}

@misc{roslyn_github,
  author    = {{Microsoft}},
  title     = {Roslyn: The {.NET} Compiler Platform},
  year      = {2014},
  url       = {https://github.com/dotnet/roslyn},
  note      = {Accessed: 2025-05-02}
}

@misc{typescript_compiler_api,
  author    = {{Microsoft}},
  title     = {Using the {TypeScript} Compiler {API}},
  year      = {2015},
  url       = {https://github.com/microsoft/TypeScript/wiki/Using-the-Compiler-API},
  note      = {Accessed: 2025-05-02}
}

@inproceedings{zhang2026round,
  title     = {Round-trip Engineering for Tactical {DDD}: A Constraint-Based Vision for the Masses},
  author    = {Zhang, Weixing and Herb, Mario and Armbruster, Martin and Jiang, Bowen and Vielsack, Marcel and Koziolek, Anne},
  booktitle = {Companion Proceedings of the 34th ACM Joint European Software Engineering Conference and Symposium on the Foundations of Software Engineering},
  series    = {FSE Companion '26},
  pages     = {5},
  year      = {2026},
  month     = jul,
  address   = {Montreal, QC, Canada},
  publisher = {ACM},
  doi       = {10.1145/3803437.3805566},
  isbn      = {979-8-4007-2636-1}
}

@article{zhong2024domain,
  author       = {Chenxing Zhong and
                  Shanshan Li and
                  Huang Huang and
                  Xiaodong Liu and
                  Zhikun Chen and
                  Yi Zhang and
                  He Zhang},
  title        = {Domain-Driven Design for Microservices: An Evidence-Based Investigation},
  journal      = {{IEEE} Trans. Software Eng.},
  volume       = {50},
  number       = {6},
  pages        = {1425--1449},
  year         = {2024},
  url          = {https://doi.org/10.1109/TSE.2024.3385835},
  doi          = {10.1109/TSE.2024.3385835},
  bibsource    = {dblp computer science bibliography, https://dblp.org}
}

@book{evans2004domain,
  title={Domain-driven design: tackling complexity in the heart of software},
  author={Evans, Eric},
  year={2004},
  publisher={Addison-Wesley Professional}
}

@book{vernon2013implementing,
  title={Implementing domain-driven design},
  author={Vernon, Vaughn},
  year={2013},
  publisher={Addison-Wesley}
}

@article{zhang2026development,
  title={Development and evolution of Xtext-based DSLs on GitHub: an empirical investigation},
  author={Zhang, Weixing and Str{\"u}ber, Daniel and Hebig, Regina},
  journal={Empirical Software Engineering},
  volume={31},
  number={3},
  pages={48},
  year={2026},
  publisher={Springer}
}

@inproceedings{zhang2024tales,
  title={Tales from 1002 repositories: Development and evolution of xtext-based dsls on github},
  author={Zhang, Weixing and Struber, Daniel},
  booktitle={2024 50th Euromicro Conference on Software Engineering and Advanced Applications (SEAA)},
  pages={172--179},
  year={2024},
  organization={IEEE}
}

@article{li2022understanding,
  title={Understanding software architecture erosion: A systematic mapping study},
  author={Li, Ruiyin and Liang, Peng and Soliman, Mohamed and Avgeriou, Paris},
  journal={Journal of Software: Evolution and Process},
  volume={34},
  number={3},
  pages={e2423},
  year={2022},
  publisher={Wiley Online Library}
}

@article{maffort2016mining,
  title={Mining architectural violations from version history},
  author={Maffort, Cristiano and Valente, Marco Tulio and Terra, Ricardo and Bigonha, Mariza and Anquetil, Nicolas and Hora, Andr{\'e}},
  journal={Empirical Software Engineering},
  volume={21},
  number={3},
  pages={854--895},
  year={2016},
  publisher={Springer}
}

@article{mumtaz2021systematic,
  title={A systematic mapping study on architectural smells detection},
  author={Mumtaz, Haris and Singh, Paramvir and Blincoe, Kelly},
  journal={Journal of Systems and Software},
  volume={173},
  pages={110885},
  year={2021},
  publisher={Elsevier}
}

@inproceedings{kapferer2020domain,
  title={Domain-driven architecture modeling and rapid prototyping with context mapper},
  author={Kapferer, Stefan and Zimmermann, Olaf},
  booktitle={International Conference on Model-Driven Engineering and Software Development},
  pages={250--272},
  year={2020},
  organization={Springer}
}

@misc{archunit,
  author       = {{TNG Technology Consulting GmbH}},
  title        = {{ArchUnit}: A {Java} Architecture Test Library},
  year         = {2017},
  howpublished = {\url{https://github.com/TNG/ArchUnit}},
  note         = {accessed: 2026-06-07}
}

@inproceedings{pietri2020forking,
  title={Forking without clicking: on how to identify software repository forks},
  author={Pietri, Antoine and Rousseau, Guillaume and Zacchiroli, Stefano},
  booktitle={Proceedings of the 17th international conference on mining software repositories},
  pages={277--287},
  year={2020}
}

@inproceedings{zhang2023manual,
  title={Manual abstraction in the wild: A multiple-case study on oss systems’ class diagrams and implementations},
  author={Zhang, Wenli and Zhang, Weixing and Str{\"u}ber, Daniel and Hebig, Regina},
  booktitle={2023 ACM/IEEE 26th International Conference on Model Driven Engineering Languages and Systems (MODELS)},
  pages={36--46},
  year={2023},
  organization={IEEE}
}

@article{zhang2025empirical,
  author       = {Wenli Zhang and
                  Weixing Zhang and
                  Daniel Str{\"{u}}ber and
                  Regina Hebig},
  title        = {An empirical study of manual abstraction between class diagrams and
                  code of open-source systems},
  journal      = {Softw. Syst. Model.},
  volume       = {24},
  number       = {6},
  pages        = {1797--1823},
  year         = {2025},
  url          = {https://doi.org/10.1007/s10270-025-01289-y},
  doi          = {10.1007/S10270-025-01289-Y},
  bibsource    = {dblp computer science bibliography, https://dblp.org}
}

\end{document}